\documentclass[12pt]{article}
\begin{document}
\date{Today}
\title{{\bf{\Large Pauli equation on noncommutative plane and the Seiberg-Witten map }}}

\author{
{\bf {\normalsize Aslam Halder}$^{a}
$\thanks{aslamhalder.phy@gmail.com}},
{\bf {\normalsize Sunandan Gangopadhyay}$^{b,c}$\thanks{sunandan.gangopadhyay@gmail.com, sunandan@iucaa.ernet.in}}\\[0.2cm]
$^{a}$ {\normalsize Kolorah H.A.W Institution, Kolorah, Howrah-711411, India}\\[0.2cm]
$^{b}$ {\normalsize Department of Physics, West Bengal State University,
Barasat, Kolkata 700126, India   }\\[0.2cm]
$^{c}${\normalsize Visiting Associate in Inter University Centre for Astronomy $\&$ Astrophysics (IUCAA),}\\
{\normalsize Pune, India}\\[0.3cm]
}
\date{}

\maketitle

\begin{abstract}
\noindent
We study the Pauli equation in noncommutative two dimensional plane which exhibits the supersymmetry algebra when the gyro-magnetic ratio is $2$. The significance of the Seiberg-Witten map in this context is discussed and its effect in the problem is incorporated to all orders in $\theta$. We map the noncommutative problem to an equivalent commutative problem by using a set of generalised Bopp-shift transformations containing a scaling parameter. The energy spectrum of the noncommutative Pauli Hamiltonian is obtained and found to be $\theta$ corrected which is valid to all orders in $\theta$.   
\end{abstract}
\vskip 1cm

\section{Introduction  }
Motivated by string theory  and also from the attempts to understand the renormalization program \cite{sw}, the discovery of noncommutative (NC) geometry has allowed the exploration of new directions in theoretical physics, namely, NC quantum mechanics \cite{duval}-\cite{sgthesis} and NC quantum field theory \cite{szabo}-\cite{carroll}. This also has drawn considerable attention to the phenomenological implications \cite{roiban}-\cite{carl} of the corresponding theories. Along with the role of noncommutativity in quantum mechanics, some other quantum effects like quantum entanglement \cite{bastos} and the collapse of the wave function \cite{bern} have also been studied in NC space.

On the other hand supersymmetry (SUSY) \cite{cooper} plays an important role in ordinary quantum mechanics. In commutative case, supersymmetric quantum mechanics is a well developed subject which has led to the understanding of many subtle issues of supersymmetric field theories in a much simpler way. As a matter of fact NC supersymmetric quantum mechanics have also been studied much in the literature in one dimension and higher dimensions \cite{hari}-\cite{scholtz}.
The aspect of NC $U(1)_{*}$ gauge invariance is an important issue in this context and had been considered in \cite{hari}. This issue is incorporated by the Seiberg-Witten (SW) map \cite{sw} which transforms NC $U(1)_{*}$ gauge system into the usual $U(1)$ gauge system in the commutative space by preserving the gauge invariance and the physics. This map is obtained by demanding that the ordinary gauge potentials, which are connected by a gauge transformation, are mapped to NC potentials which are connected by the corresponding NC gauge transformation.

In this paper, our aim is to study the Pauli equation in a two dimensional NC plane. The problem basically involves a free charged particle moving in a two dimensional NC plane with a constant background magnetic field perpendicular to the plane. The corresponding equation of motion of the particle is known as the NC Pauli equation and it is known to exhibit supersymmetry when the gyro-magnetic ratio $g = 2$. Our work incorporates the SW map to all orders in $\theta$ which is in contrast with \cite{hari} where the result is valid only up to first order in $\theta$. Further, we also use the generalised Bopp-shift transformations \cite{ah} to map the NC variables to their corresponding commutative variables. This once again differs from the analysis in \cite{hari} where Bopp-shift transformations have not been made.  In particular, we compute the energy spectrum of the NC Pauli equation and observe that the spectrum contains NC corrections. The effect of the NC parameter is also valid up to all orders in $\theta$.

This article is organised as follows. In section 2, we present a general analysis of the problem of a charged particle moving in NC plane in the presence of a constant background magnetic field and also the necessary mapping which relates the NC and the commutative sets of variables. We also discuss the SW map in this section. In section 3, we apply this general construction to obtain the energy spectrum of the NC Pauli equation. The final section is devoted to concluding remarks.

\section{General considerations}
In this section we consider a nonrelativistic charged particle of mass $m$ moving in a two dimensional NC plane in the presence of a background magnetic field $\vec{B}(\theta)=\bar{B}(\theta)\hat{k}$ perpendicular to the plane and a potential $\hat{V}$. Here $\theta$ is the spatial NC parameter and is antisymmetric in the indices $i,j$ as $\theta^{ij}=\theta \epsilon^{ij}$, where $\epsilon^{ij}=-\epsilon^{ji}, (\epsilon^{12}=1)$.
In general the NC algebra satisfied by the operators $\left(\hat{x}_i \, , \, \hat{p}_i\right)$ follow
\begin{eqnarray}
\label{e420}
\left[\hat{x}_{i},\hat{x}_{j}\right]=i\theta_{ij} = i\theta \epsilon_{ij} ~ ; \quad \left[\hat{x}_{i},\hat{p}_{j}\right] = i\hbar\delta_{ij} ~ ;\quad \left[\hat{p}_{i},\hat{p}_{j}\right]=0.
\end{eqnarray}
The standard approach in the literature to deal with such problems is to form an equivalent commutative description of the NC theory by employing some transformation which relates the NC operators $\hat{x}_{i}$, $\hat{p}_{i}$ to ordinary commutative operators $x_{i}$, $p_{i}$ satisfying the usual Heisenberg algebra 
\begin{eqnarray}
\left[x_{i}, p_{j}\right]=i\hbar \delta_{ij}~; \quad \left[x_{i} \, , \, x_{j}\right]=0= \left[p_{i},p_{j}\right].
\label{cAlgebra}
\end{eqnarray} 
The Hamiltonian of the system in this NC space reads ($\hbar=c=e=1$) 
\begin{eqnarray}
\label{e101}
\hat{H}=\frac{1}{2m}[(\vec{p}-\hat{\vec{A}})\star(\vec{p}-\hat{\vec{A}})]+\hat{V}.
\end{eqnarray}
The Schr\"{o}dinger equation of motion of the particle in NC space is therefore written as
\begin{eqnarray}
\label{e502}
i\frac{\partial \hat{\psi}(\hat{x},t)}{\partial t}=\hat{H}\star\hat{\psi}(\hat{x},t)=\hat{H}_{BS}(\theta)\hat{\psi}(\theta).
\end{eqnarray}
In the above equation first we have used star product \cite{mezin} and then the star product has been replaced by the Bopp-shift, defined in the usual way
\cite{mezin} 
\begin{eqnarray}
\label{e103}
(\hat{f}\star\hat{g})(x)=\hat{f}(x-\frac{\theta}{2}\epsilon^{ij}p_{j})\hat{g}(x).
\end{eqnarray} 
Note that $\hat{H}_{BS}$ is the Bopp-shifted Hamiltonian which is in terms of commutative variables, however $\hat{\psi}$ appearing in the eq.(\ref{e502}) is still noncommutative. In our subsequent discussion we shall use the generalised Bopp-shift \cite{ah} which comes from the consideration that the NC variables and commutative variables are related by the following set of equations
\begin{eqnarray}
\label{e74}
\hat{x}_{i}=a_{ij}x_{j}+b_{ij}p_{j}
\end{eqnarray}
\begin{eqnarray}
\label{e75}
\hat{p}_{i}=c_{ij}x_{j}+d_{ij}p_{j}
\end{eqnarray}
where $a$, $b$, $c$ and $d$ are $2\times 2$ transformation matrices. Imposing the NC Heisenberg algebra (\ref{e420}) and the usual Heisenberg algebra (\ref{cAlgebra}) on the above set of equations leads to the following set of conditions on $a_{ij}$, $b_{ij}$, $c_{ij}$, $d_{ij}$
\begin{eqnarray}
\label{e76}
ad^T-bc^T=1
\end{eqnarray}
\begin{eqnarray}
\label{e77}
ab^T-ba^T=\frac{\theta}{\hbar}
\end{eqnarray}
\begin{eqnarray}
\label{e78}
cd^T-dc^T=0.
\end{eqnarray}
To proceed further, we assume $a_{ij}=\alpha\delta_{ij}$ , $d_{ij}=\beta\delta_{ij}$, where $\alpha$ and $\beta$ are two scaling constants . 
With these assumptions, eq.(s) (\ref{e77}) and (\ref{e78}) give the solutions for the matrices $b$ and $c$ as
\begin{eqnarray}
\label{e79}
b_{ij}=-\frac{1}{2\alpha\hbar}\theta_{ij}
\end{eqnarray}
\begin{eqnarray}
\label{e80}
c_{ij}=0.
\end{eqnarray}
Now substituting these in eq.(\ref{e76}) we get
\begin{eqnarray}
\label{e8043}
\alpha\beta=1.
\end{eqnarray}
Hence the generalized Bopp-shift transformations read
\begin{eqnarray}
\label{e86}
\hat{x}_{i}  =   \alpha\left(x_{i}-\frac{1}{2\hbar\alpha^2}\theta_{ij}p_{j}\right)
\end{eqnarray}
\begin{eqnarray}
\label{e8zz} 
\hat{p}_{i}  = \frac{1}{\alpha}p_{i}.
\end{eqnarray}
With the choice $\alpha=1$, the above set of transformations reduce to the well known Bopp-shift transformations \cite{momentum_NC2} 
\begin{eqnarray}
\label{e6}
\hat{x}_{i}  =   x_{i}-\frac{1}{2\hbar}\theta_{ij}p_{j}
\end{eqnarray} 
\begin{eqnarray}
\label{e6a} 
\hat{p}_{i}  =  p_{i}.
\end{eqnarray}
To consider the $U(1)_{\star}$ gauge invariance, the SW map must be taken into account. It is a map from the NC space to commutative space which preserves the gauge invariance and the physics. To the lowest order in $\theta$, the SW maps for $\hat{\psi}$ and $\hat{A}_{k}$ read \cite{sw}, \cite{bcsgas}
\begin{eqnarray}
\label{e159}
\hat{\psi}=\psi-\frac{1}{2}\theta \epsilon^{ij}A_{i}\partial_{j}\psi
\end{eqnarray}
\begin{eqnarray}
\label{e163}
\hat{A_{k}}=A_{k}-\frac{1}{2}\theta \epsilon^{ij}A_{i}(\partial_{j}A_{k}+F_{jk}).
\end{eqnarray}
Before delving into the analysis further, we first choose a gauge for the vector potentials $\hat{A}_{i}$.
In the present discussion we take the gauge-choice in analogy with the symmetric gauge of commutative gauge theory, namely,
\begin{eqnarray}
\label{e727}
\hat{A}_{i}=-\frac{\bar{B}(\theta)}{2}\epsilon_{ij}\hat{x}^j.
\end{eqnarray}
Note that $\bar{B}(\theta)$ should not be identified with the NC magnetic field $\hat{B}$ since it has a Moyal bracket term $[\hat{A}_{1},\hat{A}_{2}]_{\star}$.

It should be noted that this gauge choice is different from that in \cite{hari} where the magnetic field involved differs from that considered in this paper. Further the coordinates appearing in this gauge choice are NC variables which is in contrast to that considered in \cite{hari}.
The NC magnetic field is given by
\begin{eqnarray}
\label{e164}
\hat{B}=\hat{F}_{12}=\partial_{1}\hat{A}_{2}-\partial_{2}\hat{A}_{1}-i[\hat{A}_{1},\hat{A}_{2}]_{\star}~.
\end{eqnarray}
Substituting the form of the vector potential from eq.(\ref{e727}) in eq.(\ref{e164}) yields the following expression for the NC magnetic field $\hat{B}$ in terms of $\bar{B}$
\begin{eqnarray}
\label{e280}
\hat{B}=\hat{F}_{12}=\bar{B}\left(1+\frac{\theta \bar{B}}{4}\right).
\end{eqnarray}
We shall now fix $\bar{B}(\theta)$ by comparing the above expression for the NC magnetic field with \cite{sw}
\begin{eqnarray}
\label{e290a}
\hat{B}=\frac{1}{1-\theta B}B.
\end{eqnarray}
This yields
\begin{eqnarray}
\label{e27}
\bar{B}(\theta)=\frac{2}{\theta}[(1-\theta B)^{-1/2}-1].
\end{eqnarray}
In the next section, we shall use the generalised Bopp-shift transformations (\ref{e86}) and the expression for $\bar{B}({\theta})$ to analyse the Pauli equation on the two dimensional NC plane.

\section{Pauli equation in noncommutative plane}
With the formalism discussed in the previous section in place, we shall now consider the problem of a nonrelativistic free charged particle of mass $m$ moving in a NC plane in the presence of a constant background magnetic field perpendicular to the plane. The Hamiltonian describing this system in NC space reads
\begin{eqnarray}
\label{e1}
\hat{H}=\frac{1}{2m}[-(\partial_{i}+\hat{A}_{i})^2+\frac{g}{2}\hat{B}\sigma_{3}]=\left(
\begin{array}{cc}
\hat{H}_2 & 0 \\
0 & \hat{H}_1 \\
\end{array}
\right)
\end{eqnarray}
where
\begin{eqnarray}
\label{e1097}
\hat{H}_1=-\frac{1}{2m}[(\partial_{i}+i\hat{A}_{i})^2+\frac{g}{2}\hat{B}]
\end{eqnarray}
\begin{eqnarray}
\label{e1096}
\hat{H}_2=\frac{1}{2m}[-(\partial_{i}+i\hat{A}_{i})^2+\frac{g}{2}\hat{B}].
\end{eqnarray}
It is a well known fact that in commutative space this problem have supersymmetry when the gyromagnetic ratio $g=2$. Interestingly, the above problem also satisfies the NC SUSY algebra for $g=2$. This can be easily seen by defining the two NC supercharges in terms of the NC variables as
\begin{eqnarray}
\label{e173}
\hat{Q}_{1}=\frac{1}{\sqrt{2m}}[i(\partial_{y}+i\hat{A}_{y})\sigma_{1}-i(\partial_{x}+i\hat{A}_{x})\sigma_{2}]
\end{eqnarray}
\begin{eqnarray}
\label{e174}
\hat{Q}_{2}=\frac{1}{\sqrt{2m}}[-i(\partial_{x}+i\hat{A}_{x})\sigma_{1}-i(\partial_{y}+i\hat{A}_{y})\sigma_{2}]
\end{eqnarray}
where $\sigma_{i}$ are the Pauli matrices. Here the two vector potentials $\hat{A}_{x}$ and $\hat{A}_{y}$ are real and act as superpotentials. In terms of these, the NC complex supercharge is defined to be
\begin{eqnarray}
\label{e175}
\hat{Q}=-\frac{i}{2}(\hat{Q}_{1}-i\hat{Q}_{2}).
\end{eqnarray}
Now it is trivial to check that the above complex supercharge satisfy the following NC SUSY algebra
\begin{eqnarray}
\label{e176}
\{\hat{Q},\hat{Q}\}_{\star}=0 ~; \quad \{\hat{Q},\hat{Q}^{\dagger}\}_{\star}=\hat{H} ~; \quad [\hat{H},\hat{Q}]_{\star}=0=[\hat{H},\hat{Q}^{\dagger}]_{\star}
\end{eqnarray}
where $\hat{H}$ is given in eq.(\ref{e1}) with $g=2$.

We shall now proceed to compute the energy spectrum of the NC Pauli equation. To do so we use the gauge choice for the vector potentials $\hat{A}_{i}$ defined in eq.(\ref{e727}).
In this gauge the Hamiltonian (\ref{e1}) is found to be
\begin{eqnarray}
\label{e2}
\hat{H}=\frac{1}{2m}\left(-(\partial_{x}-\frac{i\bar{B}}{2}\hat{y})^2-(\partial_{y}+\frac{i\bar{B}}{2}\hat{x})^2+\hat{B}\sigma_{3}\right).
\end{eqnarray}
Substituting the generalized transformations (\ref{e86}) in the above equation, we get an equivalent commutative Hamiltonian in terms of the commutative variables (operators) which describes the original system defined over the NC plane 
\begin{eqnarray}
\label{e7}
\hat{H} & = &\frac{1}{2m} \left(a^{2}p_{i}{}^2+b^{2} x_{i}{}^2 + 2 a b \epsilon_{kl}x_{k} p_{l} \right )+\frac{1}{2m}\hat{B}\sigma_{3}\\
a&=&\frac{1}{\alpha}\left(1-\frac{\theta\bar{B}}{4}\right) \quad,\quad b=\frac{\bar{B}\alpha}{2}~. \nonumber
\end{eqnarray} 
It is easy to observe that the above Hamiltonian is composed of two mutually commutating parts 
\begin{eqnarray}
\label{e3}
\hat{H}=\hat{H}_{1}+\hat{H}_{2}
\end{eqnarray}
where
\begin{eqnarray}
\label{e4}
\hat{H}_{1}& = &\frac{1}{2m} \left(a^{2}p_{i}{}^2+b^{2} x_{i}{}^2 + 2 a b \epsilon_{kl}x_{k} p_{l} \right )
\end{eqnarray}
\begin{eqnarray}
\label{e5}
\hat{H}_{2}=\frac{1}{2m}\hat{B}\sigma_{3}.
\end{eqnarray}
We now want to compute the energy spectrum of the two above parts separately. For computing the spectrum of $\hat{H}_{1}$, we introduce the ladder operators involving the commutative phase-space variables (operators) $x$, $y$, $p_{x}$, $p_{y}$
\begin{eqnarray}
\label{e30a}
a_{x}=\frac{iap_{x}+bx}{\sqrt{2ab\hbar}}~; \quad a_{x}^{\dagger}=\frac{-iap_{x}+bx}{\sqrt{2ab\hbar}}
\end{eqnarray}
\begin{eqnarray}
\label{e31a}
a_{y}=\frac{iap_{y}+by}{\sqrt{2ab\hbar}}~; \quad  a_{y}^{\dagger}=\frac{-iap_{y}+by}{\sqrt{2ab\hbar}}
\end{eqnarray}
(where $a$ and $b$ are defined in eq.(\ref{e7})) which satisfy the commutation relations 
\begin{eqnarray}
\label{e30}
[a_{x},a_{x}^{\dagger}]=1=[a_{y},a_{y}^{\dagger}].
\end{eqnarray}
In terms of these ladder operators the transformed Hamiltonian (\ref{e7}) can be rewritten as
\begin{eqnarray}
\label{e345}
\hat{H}_{1}=\frac{ab\hbar}{m}[(a_{x}^{\dagger} a_{x}+a_{y}^{\dagger} a_{y}+1)+i(a_{x}a_{y}^{\dagger}-a_{x}^{\dagger} a_{y})].
\end{eqnarray}
The above Hamiltonian is not in diagonal form. To diagonalise it, we further define the pair of operators
\begin{eqnarray}
\label{e32}
a_{+}=\frac{a_{x}+ia_{y}}{\sqrt{2}}~;\quad a_{-}=\frac{a_{x}-ia_{y}}{\sqrt{2}}
\end{eqnarray}
which satisfy the following commutation relations
\begin{eqnarray}
\label{e32a} 
[a_{+},a_{+}^{\dagger}]=1=[a_{-},a_{-}^{\dagger}].
\end{eqnarray}
In terms of these the Hamiltonian (\ref{e345}) can be recast in the following diagonal form
\begin{eqnarray}
\label{e33}
\hat{H}_{1}=\frac{2ab}{m}\left(a_{-}^{\dagger}a_{-}+\frac{1}{2}\right).
\end{eqnarray}
Therefore the energy spectrum of $\hat{H}_{1}$ reads
\begin{eqnarray}
\label{e34}
E_{1}&=&\frac{2ab}{m}\left(n+\frac{1}{2}\right)\nonumber\\
&=&\left(n+\frac{1}{2}\right)\left(\frac{\bar{B}\alpha}{2}\right)\frac{1}{\alpha}\left(1-\frac{\theta\bar{B}}{4}\right)\nonumber\\
&=&\left(n+\frac{1}{2}\right)\left(\frac{1}{m\theta}\right)\left(4(1-\theta B)^{-1/2}-(1-\theta B)^{-1}-3\right); \quad n=0,1,2,...
\end{eqnarray}
The energy eigenvalue of the operator $\hat{H}_{2}$ is given by
\begin{eqnarray}
\label{e330}
E_{2}=\pm\frac{\hat{B}}{2m}=\pm\frac{1}{2m}\left(\frac{B}{1-\theta B}\right)
\end{eqnarray}
where we have used eq.(\ref{e290a}) to relate the $\hat{B}$ with $B$.
We can now combine the two solutions to obtain the total energy eigenvalue of the system as
\begin{eqnarray}
\label{e333}
E_{n}&=&E_{1}+E_{2}\nonumber\\
&=&\left(n+\frac{1}{2}\right)\left(\frac{1}{m\theta}\right)\left(4(1-\theta B)^{-1/2}-(1-\theta B)^{-1}-3\right)\pm\frac{1}{2m}\left(\frac{B}{1-\theta B}\right);~n=0,1,2,...\nonumber\\
\end{eqnarray}
The energy spectrum is thus seen to pick up NC correction and captures its effect up to all orders in $\theta$. It is clear that the energy levels are altered by spatial noncommutativity. However these are independent of the scaling parameter $\alpha$ appearing in the expression of the generalised Bopp-shift (\ref{e86}). 

The above energy spectrum (\ref{e333}) up to first order in $\theta$ reads 
\begin{eqnarray}
\label{e333}
E_{n}=\left(n+\frac{1}{2}\pm\frac{1}{2}(1+\theta B)\right)\left(\frac{B}{m}\right); \quad n=0,1,2,...
\end{eqnarray}
which is different from that obtained in \cite{hari}. However the above result reduces to the commutative result in the $\theta=0$ limit.

\section{Conclusions}
We now summarize our findings. In this paper, we have discussed the problem of a charged particle moving in a two dimensional NC plane with a constant background magnetic field perpendicular to the plane. The approach that we have adopted to map the NC problem to an equivalent commutative problem is to use a set of generalized Bopp-shift transformations. We have also incorporated the Seiberg-Witten map to obtain the commutative equivalent Hamiltonian in terms of commutative variables and NC parameter $\theta$. Importantly, the map has been introduced to all orders in $\theta$, however the construction is valid only for the case of spatial noncommutativity.  The main result in the paper is the computation of the energy spectrum of the NC Pauli problem. The result is found to get noncommutative corrections. Further, it is independent of the scaling parameter $\alpha$ which appears in the generalised Bopp-shift
transformations. The noncommutative correction to the energy spectrum is also valid up to all orders in the noncommutative parameter $\theta$.

\section*{Acknowledgements}
S.G. acknowledges the support by DST SERB under Start Up Research Grant (Young Scientist), File No.YSS/2014/000180.

\end{document}